\RequirePackage{fix-cm}
\documentclass[smallextended]{svjour3}       
\smartqed  
\usepackage{graphicx}
\usepackage{pgfplots}
\usepackage{bm}
\usepackage{amssymb}
\begin{document}

\title{A New Integer Programming Formulation of the Graphical Traveling Salesman Problem
}

\titlerunning{A New IP Formulation of Graphical TSP}        

\author{Robert D. Carr        \and
        Neil Simonetti
}


\institute{Robert D. Carr \at
	Computer Science Department, University of New Mexico \\
	Albuquerque, NM 87131 \\
	This material is based upon research supported in part by the U. S. Office of Naval Research under award number N00014-18-1-2099.\\
	\email{bobcarr@unm.edu} \\
	\and
	Neil Simonetti \at
	Business, Computer Science, and Mathematics Department, Bryn Athyn College \\
	Bryn Athyn, PA 19009-0717 \\
	\email{neil.simonetti@brynathyn.edu}
}

\date{Received: date / Accepted: date}

\maketitle

\begin{abstract}
In the Traveling Salesman Problem (TSP), a salesman wants to visit a set of cities and return home.  There is a cost $c_{ij}$ of traveling from city $i$ to city $j$, which is the same in either direction for the Symmetric TSP.  The objective is to visit each city exactly once, minimizing total travel costs.  In the Graphical TSP, a city may be visited more than once, which may be necessary on a sparse graph.  We present a new integer programming formulation for the Graphical TSP requiring only two classes of constraints that are either polynomial in number or polynomially separable, while addressing an open question proposed by Denis Naddef.
\keywords{Linear Program \and Relaxation \and TSP \and Traveling Salesman Problem \and GTSP \and Graphical Traveling Salesman Problem}
\end{abstract}

\section{Introduction}
\label{intro}
The Traveling Salesman Problem (TSP), is one of the most studied problems in combinatorial optimization \cite{surv} \cite{surv2}.  In its classic form, a salesman wants to visit each of a set of cities exactly once and return home while minimizing travel costs.  Costs of traveling between cities are stored in a matrix where entry $c_{ij}$ indicates the cost of traveling from city $i$ to city $j$.  Units may be distance, time, money, etc.

If the underlying graph for the TSP is sparse, a complete cost matrix can still be constructed by setting $c_{ij}$ equal to the shortest path between city $i$ and city $j$ for each pair of cities. However, this has the disadvantage of turning a sparse graph $G = (V,E)$ where the edge set $E$ could be of size $O(|V|)$ into a complete graph $G' = (V,E')$, where the edge set $E'$ is $O(|V|^2)$.

Ratliff and Rosenthal were the first to consider a case where the edge set is not expanded to a complete graph, but left sparse, \cite{RR}, while soon after, Fleischmann \cite{BF} and Cornu\'ejols, Fonlupt, and Naddef \cite{CFN} examined this in a more general case, the latter giving this its name: the Graphical Traveling Salesman Problem (GTSP).   As a consequence, a city may be visited more than once, since there is no guarantee the underlying graph will be Hamiltonian.  While the works of Fleischmann and Cornu\'ejols et al. focused on cutting planes and facet-defining inequalities, this paper will look at a new compact formulation that can improve on the integrality gap created when solving a linear programming relaxation of the problem.

\section{Basic Formulations}
\label{sec1}
This paper will investigate the symmetric GTSP, where the cost of traveling between two cities is the same, regardless of direction, which allows the following notation to be used:
\begin{displaymath}
\begin{array}{rl}
G = (V,E) & \mbox{: The graph $G$ with vertex set $V$ and edge set $E$.} \\
c_e & \mbox{: The cost of using edge $e$, replaces $c_{ij}$.} \\
x_e & \mbox{: The variable indicating the use of edge $e$, replaces $x_{ij}$ which  } \\
& \mbox{~~~~~is used in most general TSP formulations.} \\
\delta(v) & \mbox{: The set of edges incident to vertex $v$.} \\
\delta(S) & \mbox{: The set of edges with exactly one endpoint in vertex set $S$.} \\
x(F) & \mbox{: The sum of variables $x_e$ for all $e\in F \subset E$.}
\end{array}
\end{displaymath}

If given a formulation on a complete graph $K_n$, a formulation for a sparse graph $G$ can be created by simply setting
$x_e = 0$ for any edge e in the graph $K_n$ but not in the graph $G$.

\subsection{Symmetric TSP}
\label{sec1a}
The standard formulation for the TSP, attributed to Dantzig, Fulkerson, and Johnson \cite{DFJ}, contains constraints that guarantee the degree of each node in a solution is exactly two (degree constraints) and constraints that prevent a solution from being a collection of disconnected subtours (subtour elimination constraints).
\begin{displaymath}
\begin{array}{llll}
\mbox{minimize}&\sum\limits_{e \in E} c_e x_e \\
\mbox{subject to} 
&\sum\limits_{e\in \delta(v)} x_e = 2 & \forall v\in V \\
&\sum\limits_{e\in \delta(S)} x_e\geq 2 & \forall S\subset V. ~S \neq\emptyset  \\
&x_e\in \left\{{0,1}\right\} & \forall e \in E.
\end{array}
\end{displaymath}
When this integer program is relaxed, the integer constraints $x_e \in \left\{{0,1}\right\}$ are replaced by the boundary constraints $0\leq x_e\leq 1$.

It should also be noted that while the subtour elimination constraints are only needed for the cases where $3\leq |S|\leq \frac{|V|}{2}$, there are still exponentially many of these constraints.  Using a similar technique to Martin \cite{RKM}, which was directly applied to the TSP by Carr and Lancia \cite{CL}, these constraints can be replaced by a polynomial number of flow constraints which ensure the solution is a 2 edge-connected graph.

\subsection{Symmetric GTSP}
\label{sec1b}
This formulation for the Graphical TSP comes from Cornu\'ejols, Fonlupt, and Naddef \cite{CFN}, and differs from the formulation above by allowing the degree of a node to be any even integer, and by removing any upper bound on the variables.
\begin{displaymath}
\begin{array}{llll}
\mbox{minimize}&\sum\limits_{e \in E} c_e x_e \\
\mbox{subject to} 
&\sum\limits_{e\in \delta(v)} x_e \mbox{~is positive and even} & \forall v\in V \\
&\sum\limits_{e\in \delta(S)} x_e \geq 2 & \forall S\subset V,~S\neq\emptyset  \\
&x_e\geq 0 & \forall e \in E. \\
&x_e \mbox{~is integer} & \forall e \in E.
\end{array}
\end{displaymath}
When this program is relaxed, the integer constraints at the end are removed, and the disjunctive constraints that require the degree of each node to be any positive and even integer are effectively replaced by a lower bound of two on the degree of each node.

The disjunctive constraints for the formuation above are unusual for two reasons.  Firstly, in most mixed-integer programs, only variables are constrained to be integers, not sums of variables found in constraints.  Secondly, the sum is required not to be just integer, but an even integer.  In terms of a mixed-integer formulation, the second perculiarity could be addressed with:
\begin{displaymath}
\begin{array}{llll}
&\sum\limits_{e\in \delta(v)} \frac{x_e}{2} \in \mathbb{Z} & \forall v\in V \\
\end{array}
\end{displaymath}

To our knowledge, no other integer programming formulation for a graph theory application uses constraints of this kind.  (Even the
constraints for the common T-join problem are different than what we are proposing here, which will be discussed at the end of the paper.)

Addressing the first peculiarity, that integer and mixed-integer programs only allow integrality of variables, we could set these sums to new variables indexed on the vertices of the graph, $d_v$.
\begin{displaymath}
\begin{array}{lrlll}
&\sum\limits_{e\in \delta(v)} \frac{x_e}{2} &=& d_v & \forall v\in V \\
&d_v &\in& \mathbb{Z} & \forall v\in V \\
\end{array}
\end{displaymath}

While this approach works, it feels unsatisfying.  The addition of these $d_v$ variables is purely cosmetic.  When solving the relaxation, there is nothing preventing us from waiting until a solution is generated before defining the values of $d_v$ using the sums above.  Thus the new variables do not facilitate the addition of any new constraints, and do nothing to strengthn the LP relaxation in any way.

When solving the integer program, we can bypass the $d_v$ variables by branching with constraints based on the degree sums.  For example, if the solution from a relaxation creates a graph where node $i$ has odd degree $q$, we branch with constraints of the form:
\begin{displaymath}
\begin{array}{lrlll}
&\sum\limits_{e\in \delta(i)} x_e &\leq& q-1 & \\
\end{array}
\end{displaymath}
and
\begin{displaymath}
\begin{array}{lrlll}
&\sum\limits_{e\in \delta(i)} x_e &\geq& q+1 & \\
\end{array}
\end{displaymath}

At a conference, Denis Naddef proposed a challenge of finding a set of constraints for a mixed-integer formulation of GTSP, where integrality constraints are limited to only $x_e \in \mathbb{Z}$ \cite{Ndf}.  We will address the state of this challenge in section~\ref{mip}.

\section{New Constraints}
\label{sec2}
Cornu\'ejols et al. proved that an upper bound of two on each $x_e$ is implied if all the edge costs are positive \cite{CFN} (and also note that without this additional bound, graphs with negative weight edges would not have finite optimal solutions). This fact allows us to dissect the variables $x_e$ into two components $y_e$ and $z_e$ such that, for each edge $e\in E$:
\begin{displaymath}
\begin{array}{rll}
y_e = 1 & \mbox{if edge $e$ is used exactly once,} & y_e = 0 \mbox{~otherwise} \\
z_e = 1 & \mbox{if edge $e$ is used exactly twice,~} & z_e = 0 \mbox{~otherwise}
\end{array}
\end{displaymath}
Note that 
\begin{equation}
x_e = y_e + 2z_e
\label{xy2z}
\end{equation}

Additionally, we can add the constraint $y_e + z_e\leq 1$ for each edge $e\in E$, since using both would imply an edge being used three times in a solution. But more importantly, we now have a way to enforce even degree without using disjunctions, since only the $y_e$ variables matter in determining if the degree of a node is odd or even.

\subsection{Enforcing even degree without disjunctions}
\label{sec2a}
Since the upper bound on the $y_e$ variables is one, the following constraints will enforce even degree:
\begin{equation}
\sum\limits_{e\in F} (1-y_e) + \sum\limits_{e\in \delta(v)\setminus F} y_e \geq 1~~\forall v\in V \mbox{and $F\subset\delta(v)$ with $|F|$ odd.}
\label{iseven}
\end{equation}

This type of constraint was used by Yannakakis et al. \cite{yan} and Lancia et al. \cite{lanc} when working with the parity polytope.

Note that for integer values of $y_e$, if the $y$-degree of a node $v$ is odd, then when $F$ is the set of nodes adjacent to $v$ indicated by $y$, the expression in the left-hand side of the constraint above will be zero.  If the $y$-degree of a node $v$ is even, then for any set $F$ with $|F|$ odd, the left-hand side must be at least one.

For sparse graphs, this adds at most $O(|V|2^{\Delta-1})$ constraints, where $\Delta$ is the maximum degree in $G$.  Typical graphs from roadmaps usually have $4\leq\Delta\leq 6$, while graphs from highway maps might have $5\leq\Delta\leq 8$.  Also note that Euler's formula for planar graphs guarantees that $|E| \leq 3|V| - 6$, and so the average degree of a node in a planar graph cannot be more than six.

\begin{figure}[h]
\begin{center}
\pgfplotsset{every axis legend/.append style={draw=none, at={(1.03,0.5)},anchor=west}}
\begin{tikzpicture}
	\begin{axis}[hide axis, axis equal image, tiny, legend style={font=\normalfont}]
		\addplot [style={solid}, black] coordinates {
			(0,0) (1.5,1) (3,1) (4.5,0) (3,-1) (1.5,-1) (0,0)
		};
		\addlegendentry{$y_e = 1$ and $z_e = 0$}
		\addplot [black, style={dashed}] coordinates {
			(0,0) (4.5,0)
		};
		\addlegendentry{$y_e = 0$ and $z_e = \frac{1}{2}$}
		\addplot [only marks, black] coordinates {
			(0,0) (1.5,1) (3,1) (4.5,0) (3,-1) (1.5,-1) (1.5,0) (3,0)
		};
	\end{axis}
\end{tikzpicture}
\caption{A path where $y_e = 0$ and $z_e = \frac{1}{2}$}
\label{fig1}       
\end{center}
\end{figure}
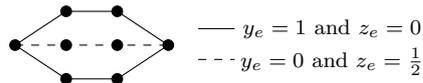

Unfortunately, in the relaxation of the linear program with these constraints, odd degree nodes can still result from allowing a path of nodes where, for each edge $e$ in the path, $y_e = 0$ and $z_e = \frac{1}{2}$.  See figure~\ref{fig1}.


\subsection{Spanning Tree Constraints}
\label{sec2d}
One method to discourage this half-$z$ path is to require the edges indicated by $\bm{y}$ and $\bm{z}$ contain a spanning tree.  This is different than demanding that $\bm{x}$ contains a spanning tree since each unit of $z_e$ contributes two units to $x_e$.  For the spanning tree constraint, each $z_e$ contributes only one unit toward a spanning tree, which means that for any node whose $y$-degree is zero, the $z$-degree must be at least one, and in the case where two nodes with $y$-degree zero are connected using an edge in the spanning tree, the $z$-degree of one of those nodes must be at least two, which effectively prevents this half-$z$ path.

Place constraints on binary variables $\bm{t}$ such that the edges where $t_e = 1$ indicate a tree that spans all nodes of $G$ (the graph must be connected and contain no cycles).  This is done by the well-known partition inequalities that will be discussed in section \ref{mip}. As with the subtour elimination constraints, Martin describes a compact set of constraints that ensure $\bm{t}$ indicates a tree (or a convex combination of trees) \cite{RKM}.  Then, add the constraint:
\begin{equation}
t_e \leq y_e + z_e,~~ \forall e \in E
\label{yztree}
\end{equation}
Only the connectedness of the graph indicated by $t_e$ is important, since the constraint only requires that $\bm{y}+\bm{z}$ dominate a spanning tree, so the constraints that would prevent cycles are unnecessary.

This constraint is valid since any tour that visits every node has within it a spanning tree that touches each node.  

\section{A New Mixed IP Formulation} 
\label{mip}
\subsection{Proving the Formulation}
\label{sec4a}
The tree constraints are sufficient, when combined with constraint (\ref{iseven}) and integrality constraints $y_e\in\left\{0,1\right\}$, to find an optimal integer solution value, making the subtour elimination constraints unnecessary.  The following new mixed integer programming formulation therefore does not include these optional constraints.  Note that all GTSP tours will satisfy the constraints in this formulation.  The notation
$\delta(V_1, ..., V_k)$ refers to the set of edges with endpoints in different vertex sets.

\begin{displaymath}
\begin{array}{lcll}
\mbox{minimize}&\sum\limits_{e \in E} c_e x_e \\
\mbox{subject to}
&x_e = y_e + 2z_e & \forall e\in E & (4.1) \\
&\sum\limits_{e\in F} 1-y_e + \sum\limits_{e\in \delta(v)\setminus F} y_e \geq 1 & \forall v\in V \mbox{and $F\subset\delta(v)$ with $|F|$ odd} & (4.2) \\
&\sum\limits_{e\in \delta(V_1, ..., V_k)} t_e \geq k-1 & \forall \mbox{~partitions~} V_1, ..., V_k \mbox{~of~} V & (4.3) \\
&t_e \leq y_e + z_e \leq 1 & \forall e\in E & (4.4) \\
&0\leq t_e \leq 1 & \forall e \in E & (4.5) \\
&0\leq z_e \leq 1 & \forall e \in E & (4.6) \\
&y_e\in \left\{0,1\right\}& \forall e \in E & (4.7) \\
\end{array}
\end{displaymath}

\begin{theorem}
Given a MIP solution $(\bm{y^*}, \bm{z^*})$ to the GTSP formulation above, then $\bm{x^*} = \bm{y^*} + 2\bm{z^*}$ will indicate an edge set that is an Euler tour, or a convex combination of Euler tours.
\end{theorem}

$Proof$.  It should be noted that it is sufficient for the MIP solultion to dominate an Euler tour (or a convex combination of them), since if there is some edge $e$ where $x_e^*$ is larger than necessary by $\epsilon$ units (for some $0 < \epsilon \leq x^*_e \leq 2$), one can add edge $e$ twice to a collection of Euler tours with total weight $\frac{\epsilon}{2}$.

Let $(\bm{y^*}, \bm{z^*})$ be a feasible solution to the GTSP formulation specified.
By constraints (4.3) and (4.4), we know that $\bm{y^*}+\bm{z^*}$ dominates a convex combination of spanning trees, thus we have
$\bm{y^*}+\bm{z^*} \geq \sum_i \lambda_i \bm{T^i}$, where
each $\bm{T^i}$ is an edge incidence vector of a spanning tree.
Define $\bm{R^i}$ by $R_e^i = 1$ if both $T_e^i = 1$ and $y_e^* = 0$, and $R_e^i = 0$ otherwise.  So $\bm{R^i}$ becomes the remnant of tree
$\bm{T^i}$ when edges indicated by $\bm{y^*}$ are removed.  Since $\bm{y^*}$ only contains integer values, the constraint
$y_e + z_e \leq 1$ guarantees that $z_e = 0$ whenever $y_e = 1$, and guarantees $y_e = 0$ whenever $z_e > 0$.  This means that for all edges where
$y_e = 0$, we have $\bm{z^*} \geq \sum_i \lambda_i \bm{T^i} = \sum_i \lambda_i \bm{R^i}$. Since $\bm{R^i}$ is the result of removing edges from $\bm{T^i}$ where $y_e = 1$, we are guaranteed that $R_e^i = 0$ for every $i$ where $y_e = 1$, and thus 
$\bm{z^*} \geq \sum_i \lambda_i \bm{R^i}$ over all edges.

Hence, $\bm{x^*} = \bm{y^*} + 2\bm{z^*} \geq \bm{y^*} + 2\sum_i \lambda_i \bm{R^i} = \sum_i \lambda_i (\bm{y^*} +2\bm{R^i})$, where for each $i$, $\bm{y^*} +2\bm{R^i}$ is an Euler tour, because constraint (4.2) ensures the graph indicated by $\bm{y^*} +2\bm{R^i}$ will have even degree at every node, and constraints (4.3) and (4.4) ensure the graph indicated by $\bm{y^*} +\bm{R^i}$, and thus $\bm{y^*} +2\bm{R^i}$, is connected. \qed

Constraints (4.3) are exponential in number, but these can also be reduced to a compact set of constraints using the techniques from Martin \cite{RKM}.  The constraints we used are below, and require a model using directed edges to regulate flow variables $\bm{\phi}$. 
Assume $V = \{1, 2, ..., n\}$ and designate city $n$ as the home city.

In Martin’s formulation, we use directed flow variables $\bm{\phi^k}$ that carry one unit of flow from any node with index higher than $k$ to node $k$ and supported by the values $\overrightarrow{t}_{ij}$ as edge capacities.
From any feasible integral solution (directed spanning tree), it is not hard to derive such a set of unit flows by directing the tree from the home node $n$.
For this flow, we can now set flow going into any node$j$ with $j>k$ to zero in flow problem $\bm{\phi^k}$ , and flow balancing constraints among the
nodes numbered $k+1$ or higher are also unnecessary.  Finally, $t_e = \overrightarrow{t}_{ij} + \overrightarrow{t}_{ji}$ to create variables for the undirected spanning tree.


\begin{displaymath} 
\begin{array}{rlll}
\phi^k_{i,j} &=& 0 & \forall j\in V, k\in V\setminus\{n\} \mbox{~with~} j > k,\{i,j\}\in E \\[4pt]
\phi^k_{k,i} &=& 0 & \forall k\in V\setminus\{n\},\{i,k\}\in E \\[4pt]
\sum\limits_{i\in \delta(k)} \phi^k_{i,k} &=& 1 & \forall k\in V\setminus\{n\} \\
\sum\limits_{i\in \delta(j)} \phi^k_{i,j} - \sum\limits_{i\in \delta(j)} \phi^k_{j,i} &=& 0 & \forall j\in V, k\in V\setminus\{n\} \mbox{~with~} j < k \\
0\leq\phi^k_{i,j} &\leq& \overrightarrow{t}_{ij} & \forall k\in V\setminus\{n\}, \{i,j\} \in E \\[4pt]
t_e & = & \overrightarrow{t}_{ij} + \overrightarrow{t}_{ji}  & \forall e=\{i,j\} \in E \\[4pt]
\sum\limits_{e\in E} t_e&\leq&n-1 \\ 
t_e&\leq& y_e + z_e & \forall e\in E 
\end{array}
\end{displaymath}

Constraints (4.2) are exponential in $\Delta$, the maximum degree of the graph, which is not a concern if the graph is sparse, leading to a compact formulation.
If the graph is not sparse, identifying when a constraint from the set (4.2) is violated, a process called separation, can be done quickly and efficiently, even if the solution is from a relaxation and thus contains fractional values for some $y_e$ variables. 

\begin{theorem}
Given a solution to the relaxation of the GTSP formulation above without constraints (4.2), if a constraint from (4.2) is violated, it can be found in $O(|V|^2)$ time.
\end{theorem}

$Proof$.  For each node $v\in V$, minimize the left-hand side of constraint (4.2) over all possible sets $F\subset\delta(v)$ ($|F|$ even or odd), by placing edges with $y_e > \frac{1}{2}$ in $F$ and leaving edges with $y_e < \frac{1}{2}$ for $\delta(v)\setminus F$.  Edges with $y_e = \frac{1}{2}$ could go in either set.
\begin{itemize}
\item If this minimum is not less than 1, no constraint from (4.2) will be violated for this node.
\item If the minimum is less than 1, and $|F|$ is odd, this is a violated constraint from (4.2).
\item If the minimum is less than 1, and $|F|$ is even, find the edge $e$ where $|y_e -  \frac{1}{2}|$ is smallest.  Then flip the status of the membership of edge $e$ in $F$.  This will create the minimum left-hand side over all sets $F$ with $|F|$ odd.
\end{itemize}

For each node, this requires summing or searching items indexed by $\delta(v)$ a constant number of times, and since $|\delta(v)| < |V|$ this requires $O(|V|^2)$ time.\qed

\subsection{Addressing the Naddef Challenge}
\label{sec4b}

We would have preferred to simply require the values in $\bm{x}$ to be integer and allow $\bm{y}$ and $\bm{z}$ to hold fractional values, which addresses the challenge that Denis Naddef proposed \cite{Ndf}.  He wished to know if one could find a simple formulation for the GTSP that finds optimal solutions by only requiring integrality of the decision variables $\bm{x^*}$, and nothing else.  But this cannot be done (for polynomially-sized or polynomially-separable classes of inequalities unless $P=NP$), which can be seen by the folowing theorem.

\begin{theorem}
Let $G$ be a 3-regular graph, and let $G'$ be the result of adding one vertex to the middle of each edge in $G$.  Consider a
solution $\bm{x^*}$, where $x^*_e = 1$ for each edge $e \in G'$.  Then $\bm{x^*}$ is in the GTSP polytope iff $G$ is Hamiltonian.
\end{theorem}
\label{hammy}
In this proof, define $x^*(S) = \sum_{e \in S} x^*_e$.

$Proof$.  If $G$ is Hamiltonian, let $P$ be the set of edges in a Hamilton cycle of $G$, and let $P'$ be the set of corresponding edges in the graph $G'$.  Note that in $G'$ every degree-two node is adjacent to two degree-three nodes, and that the cycle $P'$ reaches every degree-three node in $G'$.  One GTSP tour in $G'$ can be created by adding an edge of weight two on exactly one of the two edges adjacent to each degree-two node in $G'$ but not used in $P'$.  The other GTSP tour can be created by adding an edge of weight two on the edges not chosen by the first tour.  The convex combination of each of these tours with weight $\frac{1}{2}$ will create a solution where $x^*_e = 1$ for each edge $e \in G'$

Now suppose we have a solution $\bm{x^*}$, where $x^*_e = 1$ for each edge $e \in G'$ and $\bm{x^*}$ is in the GTSP polytope.  Express $\bm{x^*} = \sum_k \lambda_k \bm{\chi^k}$ as a convex combination of GTSP tours.  Consider any degree-two vertex $v$ in $G'$.  Since $v$ has degree two, $\bm{x^*}(\delta(v)) = 2$.  Also $\bm{\chi^k}(\delta(v)) \geq 2$ must be true for any GTSP tour $\bm{\chi^k}$, and so, by the convex combination, it must be that $\bm{\chi^k}(\delta(v)) = 2$ for each $\bm{\chi^k}$.  If the neighbors of $v$ are nodes $i$ and $j$, then $\bm{\chi^k}(\delta(v)) = 2$ implies either $\chi^k_{i,v}=1$ and $\chi^k_{j,v}=1$, or $\chi^k_{i,v}=2$ and $\chi^k_{j,v}=0$, or $\chi^k_{i,v}=0$ and $\chi^k_{j,v}=2$.  The edges of weight one in $\bm{\chi^k}$ form disjoint cycles, so pick one such cycle $C$, and let $S_1$ be the set of vertices in $C$.  (If there are no edges of weight one in $\bm{\chi^k}$, let $S_1$ be a set containing any single degree three vertex.)  Let $S_2$ be the set of degree two vertices $v$ such that $\chi^k_{i,v}=2$ for some $i \in S_1$.  Notice that $\bm{\chi^k}(\delta(S_1 \cup S_2)) = 0$, since the degree of any node, $v\in S_2$ is exactly two in $\bm{\chi^k}$, and the edge connecting $v$ to $S_1$ has weight two. 
$\bm{\chi^k}$ is a tour and thus must be connected, which is only possible if $S_1 \cup S_2$ is the entire vertex set of $G'$, and therefore the cycle $C$ must visit every degree three node in $G'$.  The corresponding cycle in the graph $G$ would therefore be a Hamilton cycle. \qed

If one knew when the integer solution $\bm{x^*}$ were in the GTSP polytope, then this theorem would imply a polynomial time algorithm to determine if a 3-regular graph is Hamiltonian, which is an $NP$-complete problem.

The challenge that Naddef proposed never specifically defined what makes a formulation simple.  Certainly having all constraint sets be polynomially-sized or polynomially-separable (in terms of $n$, the number of nodes) would qualify as simple, but there may be other normal sets of constraints that could satisfy the spirit of Naddef's challenge.  One such example is Naddef's conjecture that simply using the three classes of inequalities from his 1985 paper (path, wheelbarrow, and bicycle inequalities) \cite{CFN} with integrality constraints only on the variables $\bm{x}$, would be sufficient to formulate the problem.  Since it is not known if these three classes of inequalities can be separated in polynomial time, the theorem above does not directly address this conjecture.

However, if we are given an arbitrary constraint of the form $\bm{ax} \geq b$, it can be recognized in polynomial time whether or not this constraint belongs to a particular class of inequality (path, wheelbarrow, or bicycle) and whether or not a potential solution $\bm{x^*}$ violates this constraint.
If that potential solution $\bm{x^*}$ were not in the GTSP polytope, then there would be a polynomially sized verification of the graph $G$ not being Hamiltonian, which would imply $NP$ = co-$NP$.

This would apply to any integer programming formulation with a finite number of inequality classes that contain inequalties that are normal.  In this case, we define normal to mean that the membership of any individual constraint in a class can be verified in polynomial time.

This implies Naddef's challenge cannot be completed successfully using normal inequalities, unless $NP$ = co-$NP$.  However, our formulation follows its spirit, as the integer constrained variables in our formulation $\bm{y}$ have a one-to-one correspondence to the integer constrained variables $\bm{x}$ in the challenge.

\subsection{Interesting Notes Concerning Degree Two Nodes in GTSP}
\label{sec4c}

It is difficult to solve Naddef's challenge because the integer programming formulation for the GTSP has $x_e \in \{0, 1, 2\}$, whereas most formulations for other graph theory applications simply require $x_e \in \{0, 1\}$.  In the variant of GTSP where doubled edges are disallowed, but nodes may still be visited multiple times, the formulation from above would be a solution to the Naddef challenge, since in this case, $\bm{x} = \bm{y}$ and $\bm{z} = \bm{0}$.  If there are degree 2 nodes present in the graph, then disallowing doubles edges forces the tour across a particular path, since the tour cannot visit this degree 2 node and return back along the same edge.

Assume we have an integer programming formulation for the GTSP.  Then any integer point $\bm{x^*}$ dominates a convex combination of GTSP tours, or it must violate at least one inequality from this formulation.  Therefore, using the graphs $G$ and $G'$ illustrated in Theorem 3 (where $G$ is a 3-regular graph, and $G'$ is the same graph with every edge subdivided into two with a new degree 2 node), the GTSP formulation when applied to $G'$ would certify whether the graph $G$ is Hamiltonian or non-Hamiltonian.  If the constraint classes of the formulation are normal, as defined at the end of the previous section, then this certificate can be constructed in polynomial time.

In the case where $G$ is not Hamiltonian, an integer programming formulation for the GTSP must have a violated constraint for any solution $\bm{x^*}$ where
$x^*(E') = |E'|$, where $E'$ is the edge set of $G'$.  Assuming $n = |V|$, the node set of $G$, and $n' = |V'|$, the node set of $G'$, we can determine the size of $|E'|$ by noting that every edge in $E'$ connects a degree 3 node to a degree 2 node.  Therefore, the set of degree 2 
nodes can be represented by $W = V' \setminus V$ and $|E'|$ is equal to the number of degree 2 nodes in $G'$ times two, or $2(n'-n)$.  The number of degree 2 nodes in $G'$ is the same as the number of edges in $G$, which is $\frac{3}{2}n$, so we get
$n'-n = \frac{3}{2}n$ or $n = \frac{2}{5}n'$, and $2(n'-n) = 2(n'- \frac{2}{5}n') = \frac{6}{5}n'$.  Since $x(\delta(v')) \geq 2$
for any node in $v' \in G'$, 
adding these constraints over all degree 2 nodes gives the constraint:
\begin{displaymath} 
\begin{array}{c}
\sum\limits_{v'\in W} x(\delta(v')) =  x(E') \geq \frac{6}{5}n'
\end{array}
\end{displaymath}

Since the graph is not Hamiltonian, this contraint cannot be satisfied at equality, leading to $x(E') > \frac{6}{5}n'$. Since the expression $\frac{6}{5}n'$ is equal to two times an integer quantity $(n'-n)$, $\frac{6}{5}n'$ must be an even integer.  Furthermore, for any solution to the formulation, $x(\delta(v'))$ must be positive and even for every degree 2 node $v' \in G'$, and thus $x(E')$ must also be even, so the constraint can be stated as: 
\begin{displaymath} 
\begin{array}{c}
\sum\limits_{v'\in W} x(\delta(v')) =  x(E') \geq \frac{6}{5}n' + 2
\end{array}
\end{displaymath}

But these inequalities do not make up a normal class, as defined in the previous section.  This is because the validity of this inequality relies on the certainty of $G$ being non-Hamiltonian.  We believe these inequalities can be lifted to a complete graph with no coefficient greater than 3 (we are quite sure we could do this with maximum coefficient 4).

\begin{figure}[h]
\begin{center}
\pgfplotsset{every axis legend/.append style={draw=none, at={(1.03,0.5)},anchor=west}}
\begin{tikzpicture}
	\begin{axis}[hide axis, axis equal image, tiny, legend style={font=\normalfont}]
		\addplot [style={solid}, black] coordinates {
			(7,-3.25) (2,-2) (0,0) (2,2) (7,3.5)
		};
		\addplot [style={solid}, black] coordinates {
			(0,0) (6,0)
		};
		\addplot [style={solid}, black] coordinates {
			(6,3) (7,2.5)
		};
		\addplot [style={solid}, black] coordinates {
			(6,-3) (7,-2.5)
		};
		\addplot [style={solid}, black] coordinates {
			(7,0.5) (6,0) (7,-0.5)
		};
		\addplot [only marks, black] coordinates {
			(0,0) (2,2) (4,2.5) (6,3) (2,0) (4,0) (6,0) (2,-2) (4,-2.5) (6,-3)
		};
		\addplot [domain=-pi:pi,samples=200,black,line width=0.7pt]({1.2+2*sin(deg(x))}, {3*cos(deg(x))});
		\addplot [domain=-pi:pi,samples=200,black,line width=0.7pt]({3+2*sin(deg(x))}, {2.25+0.9*cos(deg(x))});
		\addplot [domain=-pi:pi,samples=200,black,line width=0.7pt]({3+2*sin(deg(x))}, {0.9*cos(deg(x))});
		\addplot [domain=-pi:pi,samples=200,black,line width=0.7pt]({3+2*sin(deg(x))}, {-2.25+0.9*cos(deg(x))});
	\end{axis}
\end{tikzpicture}
\caption{A violated 3-tooth comb inequality}
\label{comb}       
\end{center}
\end{figure}
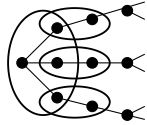

Another interesting graph involving degree 2 nodes comes from subdividing an edge twice, creating a path of three edges with two intermediate degree 2 nodes.  Given any 3-regular, 3-edge connected graph, Haddadan et al. \cite{Ravi} showed that the point $x^*$ where $x_e = 1$ for every edge in such a graph will be in the GTSP polytope, even though every node in the graph has odd degree.  Now imagine choosing any individual degree 3 node, call it $v$, and subdividing each of the incident edges twice, creating six degree 2 nodes, two along each path.  Now the solution $x^*$ where $x_e = 1$ for every edge cannot be in the GTSP polytope, since we can easily find a violated 3-tooth comb inequality by choosing $v$ and its immediate neighbors for the handle, and the pairs of adjacent degree 2 nodes as the teeth (see figure \ref{comb}).

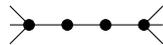
\begin{figure}[h]
\begin{center}
\pgfplotsset{every axis legend/.append style={draw=none, at={(1.03,0.5)},anchor=west}}
\begin{tikzpicture}
	\begin{axis}[hide axis, axis equal image, tiny, legend style={font=\normalfont}]
		\addplot [style={solid}, black] coordinates {
			(0,1) (1,0) (0,-1)
		};
		\addplot [style={solid}, black] coordinates {
			(8,1) (7,0) (8,-1)
		};
		\addplot [style={solid}, black] coordinates {
			(0,0) (8,0)
		};
		\addplot [only marks, black] coordinates {
			(1,0) (3,0) (5,0) (7,0)
		};
	\end{axis}
\end{tikzpicture}
\caption{A path of 3 edges connecting 2 higher-degree nodes with interior nodes of degree 2}
\label{path}       
\end{center}
\end{figure}

Furthermore, consider a solution $\bm{x^*}$ that is in the GTSP polytope for a graph, where $x_e = 1$ along each edge of a path of at least three edges connecting two higher-degree nodes with only degree 2 nodes along the interior of the path (see figure \ref{path}).
Then every GTSP tour that makes up the convex combination of tours for the solution $\bm{x^*}$ must also have $x_e = 1$ along each edge of the path.  To prove this for a path $P$ of three edges, notice that $x(P) \geq 3$ for any GTSP tour, and since $x^*(P) = 3$, we know $x(P) = 3$ for every GTSP tour in the convex combination indicated by $\bm{x^*}$.  If $x_e = 2$ for some edge in the path, then since the path has three edges, $x(P) \geq 4$, and thus could not be in the convex combination of tours indicated by $\bm{x^*}$.  Alternatively, consider a solution $\bm{x^*}$ that is in the GTSP polytope for a graph, where $x_e = 1$ along each edge of a path of only two edges connecting two higher-degree nodes with a degree 2 node in the middle of the path.  Note that an edge with $x_e = 2$ may be in one of the GTSP tours in the convex combination indicated by $\bm{x^*}$, since one tour of weight $\frac{1}{2}$ could visit one edge of the path twice, and another tour of weight $\frac{1}{2}$ could visit the other edge twice.

\section{Relaxations and Steiner Nodes} 

\subsection{Symmetric GTSP with Steiner Nodes}
\label{sec1c}
Cornu\'ejols et al. also proposed a variant of the GTSP where only a subset of nodes are visited \cite{CFN}.  As in most road networks, one may travel through many intersections that are not also destinations when traveling from one place to another.  Cornu\'ejols et al. referred to these intersection nodes as Steiner nodes.  This creates a formulation on a graph $G = (V_d\cup V_s, E)$ with $V_d\cap V_s = \emptyset$ where $V_d$ represents the set of destination nodes and $V_s$ represents the set of Steiner nodes.
\begin{displaymath}
\begin{array}{llll}
\mbox{minimize}&\sum\limits_{e \in E} c_e x_e \\
\mbox{subject to} 
&\sum\limits_{e\in \delta(v)} x_e \mbox{~is positive and even} & \forall v\in V_d \\
&\sum\limits_{e\in \delta(v)} x_e \mbox{~is even} & \forall v\in V_s \\
&\sum\limits_{e\in \delta(S)} x_e \geq 2 & \forall S\subset V \mbox{~where~} S\cap V_d\neq \emptyset, \neq V_d  \\
&x_e\geq 0 & \forall e \in E. \\
&x_e \mbox{~is integer} & \forall e \in E.
\end{array}
\end{displaymath}
Note that only sets that include destination nodes need to have corresponding cut constraints, and these can be limited to sets where the intersection is $\frac{|V_d|}{2}$ or smaller.  Again, these can be replaced by the flow constraints in the style proposed by Martin \cite{RKM}.  The constraints used in our computational results are similar to those in the multi-commodity flow formulation found by Letchford, Nasiri, and Theis \cite{LNT}.  We were able to reduce the number of variables used by Letchford, et al. by a factor of 2, by setting many variables to 0,
as we did with the flow variables in the formulation of section \ref{sec4a}.  Assume $V_d = \{1, 2, ..., d\}$ and $V_s = \{d+1, d+2, ... n\}$ and designate city $d$ as the home city.  We include $d-1$ flow problems, where each problem requires that 2 units of flow pass from nodes $S = \{k+1, k+2, ..., d\}$ to node $k$ using the values of  $\bm{x}$ as edge capacities.
Since nodes in $S$ are all sources, we can set flow into these nodes to zero, as well as setting the flow coming out of node $k$ to zero.
\begin{displaymath} 
\begin{array}{rlll}
f^k_{i,j} &=& 0 & \forall j\in V_d, k\in V_d\setminus\{d\} \mbox{~with~} j > k,\{i,j\}\in E \\[4pt]
f^k_{k,i} &=& 0 & \forall k\in V_d\setminus\{d\}, \{i,k\}\in E\\[4pt]
\sum\limits_{i\in \delta(k)} f^k_{i,k} &=& 2 & \forall k\in V_d\setminus\{d\} \\
\sum\limits_{i\in \delta(j)} f^k_{i,j} - \sum\limits_{i\in \delta(j)} f^k_{j,i} &=& 0 & \forall j\in V, k\in V_d\setminus\{d\} \mbox{~with either~} j < k \mbox{~or~} j\in V_s \\
f^k_{i,j} + f^k_{j,i}&\leq& x_e & \forall  k\in V_d\setminus\{d\}, \{i,j\}=e\in E
\end{array}
\end{displaymath}

\subsection{Preventing the Half-$z$ Path without Spanning Trees}
\label{sec2b}
While the spanning tree constaints (\ref{yztree}) of section \ref{sec2d} can prevent half-$z$ paths when integrality of $\bm{y}$ is enforced, for the computational results in the next section, better integrality gaps were obtained by using the subtour elimination
constraints plus the following, which prevents half-$z$ paths (with three or more edges) without requiring the
integrality of $\bm{y}$.
\begin{equation}
\sum\limits_{e'\in \delta(i)} x_{e'} + \sum\limits_{e'\in \delta(j)} x_{e'} - 2z_e\geq 4~~\forall e\in E
\label{bobslatest}
\end{equation}
where $i$ and $j$ are endpoints of edge $e$.

If $z_e = 1$, this constraint is the subtour elimination constraint for the set $\left\{ {i,j} \right\}$.
If $z_e = 0$, this is the sum of the degree constraints (lower bound) for nodes $i$ and $j$.
But in the middle of a path of length three or longer with edges that have $y_e = 0$ and $z_e = \frac{1}{2}$, the left side of this constraint will only add to three.

It should be noted that this constraint can only be used when both endpoints of $e$ are destination nodes, since Steiner nodes do not have a lower bound of degree 2, but could be degree zero.

\begin{figure}[h]
\begin{center}
\pgfplotsset{every axis legend/.append style={draw=none, at={(1.1,0.5)},anchor=west}}
\begin{tikzpicture}
	\begin{axis}[hide axis, axis equal image, legend style={font=\normalfont}]
		\addplot [style={solid}, black] coordinates {
			(0,10) (3,11) (9,11) (12,10) (9,9) (3,9) (0,10)
		};
		\addlegendentry{$y_e = 1$ and $z_e = 0$}
		\addplot [black, style={dashed}] coordinates {
			(0,10) (12,10)
		};
		\addlegendentry{$y_e = 0$ and $z_e = \frac{1}{2}$}
		\addplot [style={solid}, black, line width=2pt] coordinates {
			(6,6) (9,6)
		};
		\addlegendentry{$y_e = 0$ and $z_e = 1$}
		\addplot [style={solid}, black] coordinates {
			(0,6) (3,7) (9,7) (12,6) (9,5) (3,5) (0,6)
		};
		\addplot [black, style={dashed}] coordinates {
			(0,6) (12,6)
		};
		\addplot [style={solid}, black] coordinates {
			(0,2) (3,3) (9,3) (12,2) (9,1) (3,1) (0,2)
		};
		\addplot [style={solid}, black, line width=2pt] coordinates {
			(0,2) (9,2)
		};
		\addplot [only marks, black] coordinates {
			(0,10) (3,11) (6,11) (9,11) (12,10) (9,9) (6,9) (3,9) (3,10) (6,10) (9,10)
			(0,6) (3,7) (6,7) (9,7) (12,6) (9,5) (6,5) (3,5) (3,6) (6,6) (9,6)
			(0,2) (3,3) (6,3) (9,3) (12,2) (9,1) (6,1) (3,1) (3,2) (6,2) (9,2)
		};
		\addplot[scatter, only marks, nodes near coords, nodes near coords style={font=\small}, point meta=explicit symbolic, mark size=0pt] 
			table [meta=lab] {
			x      y      lab    
			6    7.8  {Objective value 12 using only constraint (\ref{iseven})} 
			6    3.8  {Objective value 13 using both constraints (\ref{iseven}) and (\ref{bobslatest})} 
			6    -0.2  {Objective value 14 for an integer solution}
		};
	\end{axis}
\end{tikzpicture}
\caption{Solutions from a 3-path configuration of four edges each}
\label{3p11}       
\end{center}
\end{figure}
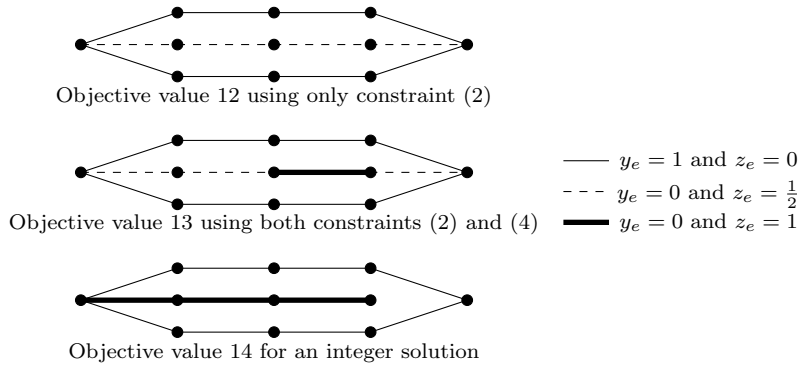

It should also be noted that if the GTSP instance is composed only of three paths of length three between two specific nodes
(see figure \ref{fig1} from section \ref{sec2a}) constraints (\ref{iseven}) from section \ref{sec2a} (those that enforce even degree) and (\ref{bobslatest}) (defined above) will be enough to close the entire integrality gap using an LP relaxation.   If the paths are all four or more edges long, this constraint will not eliminate the integrality gap, but will help.  (See figure \ref{3p11})  

\begin{figure}[h]
\begin{center}
\pgfplotsset{every axis legend/.append style={draw=none, at={(1.03,0.5)},anchor=west}}
\begin{tikzpicture}
	\begin{axis}[hide axis, axis equal image, legend style={font=\normalfont}]
		\addplot [style={solid}, black] coordinates {
			(0,10) (2,11) (12,11) (14,10) (12,9) (2,9) (0,10)
		};
		\addlegendentry{$y_e = 1$ and $z_e = 0$}
		\addplot [black, style={dashed}] coordinates {
			(0,10) (14,10)
		};
		\addlegendentry{$y_e = 0$ and $z_e = \frac{1}{2}$}
		\addplot [style={solid}, black, line width=2pt] coordinates {
			(4,6) (6,6)
		};
		\addlegendentry{$y_e = 0$ and $z_e = 1$}
		\addplot [style={solid}, black, line width=2pt] coordinates {
			(10,6) (12,6)
		};
		\addplot [style={solid}, black] coordinates {
			(0,6) (2,7) (12,7) (14,6) (12,5) (2,5) (0,6)
		};
		\addplot [black, style={dashed}] coordinates {
			(0,6) (14,6)
		};
		\addplot [style={solid}, black] coordinates {
			(0,2) (2,3) (12,3) (14,2) (12,1) (2,1) (0,2)
		};
		\addplot [style={solid}, black, line width=2pt] coordinates {
			(0,2) (12,2)
		};
		\addplot [only marks, black] coordinates {
			(0,10) (2,11) (4,11) (6,11) (8,11) (10,11) (12,11) (14,10) (12,9) (10,9) (8,9) (6,9) (4,9) (2,9)
				(2,10) (4,10) (6,10) (8,10) (10,10) (12,10)
			(0,6) (2,7) (4,7) (6,7) (8,7) (10,7) (12,7) (14,6) (12,5) (10,5) (8,5) (6,5) (4,5) (2,5) (2,6) (4,6) (6,6) (8,6) (10,6) (12,6)
			(0,2) (2,3) (4,3) (6,3) (8,3) (10,3) (12,3) (14,2) (12,1) (10,1) (8,1) (6,1) (4,1) (2,1) (2,2) (4,2) (6,2) (8,2) (10,2) (12,2)
		};
		\addplot[scatter, only marks, nodes near coords, nodes near coords style={font=\small}, point meta=explicit symbolic, mark size=0pt] 
			table [meta=lab] {
			x      y      lab    
			7    7.8  {Objective value 21 using only constraint (\ref{iseven})} 
			7    3.8  {Objective value 23 using both constraints (\ref{iseven}) and (\ref{bobslatest})} 
			7    -0.2  {Objective value 26 for an integer solution}
		};
	\end{axis}
\end{tikzpicture}
\caption{Solutions from a 3-path configuration of seven edges each}
\label{3p20}       
\end{center}
\end{figure}
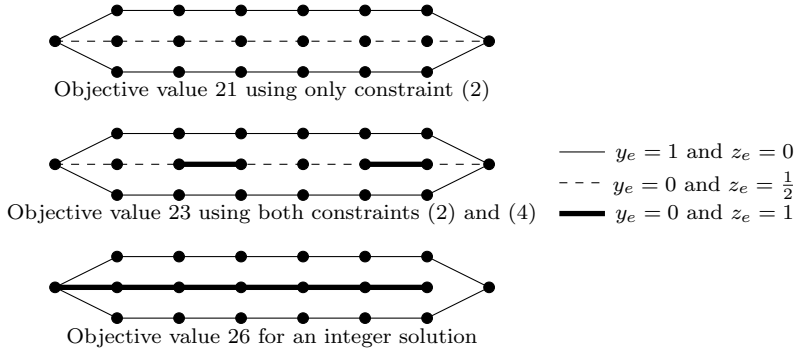

As the paths get longer, the integrality gap slowly grows. The spanning tree constraints will be useful once the paths reach a length of at least seven.  (See figure \ref{3p20}) 

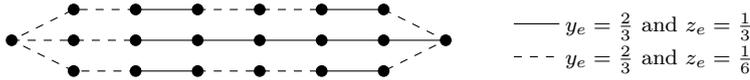
\begin{figure}[h]
\begin{center}
\pgfplotsset{every axis legend/.append style={draw=none, at={(1.03,0.5)},anchor=west}}
\begin{tikzpicture}
	\begin{axis}[hide axis, axis equal image, legend style={font=\normalfont}]
		\addplot [style={solid}, black] coordinates {
			(4,11) (6,11)
		};
		\addlegendentry{$y_e =  \frac{2}{3}$ and $z_e =  \frac{1}{3}$}
		\addplot [black, style={dashed}] coordinates {
			(4,11) (2,11) (0,10) (2,9) (4,9)
		};
		\addlegendentry{$y_e =  \frac{2}{3}$ and $z_e = \frac{1}{6}$}
		\addplot [style={solid}, black] coordinates {
			(10,11) (12,11)
		};
		\addplot [style={solid}, black] coordinates {
			(4,9) (6,9)
		};
		\addplot [style={solid}, black] coordinates {
			(10,9) (12,9)
		};
		\addplot [style={solid}, black] coordinates {
			(4,10) (14,10)
		};
		\addplot [black, style={dashed}] coordinates {
			(0,10) (4,10)
		};
		\addplot [black, style={dashed}] coordinates {
			(6,11) (10,11)
		};
		\addplot [black, style={dashed}] coordinates {
			(6,9) (10,9)
		};
		\addplot [black, style={dashed}] coordinates {
			(12,11) (14,10) (12,9)
		};
		\addplot [only marks, black] coordinates {
			(0,10) (2,11) (4,11) (6,11) (8,11) (10,11) (12,11) (14,10) (12,9) (10,9) (8,9) (6,9) (4,9) (2,9)
				(2,10) (4,10) (6,10) (8,10) (10,10) (12,10)
		};
	\end{axis}
\end{tikzpicture}
\caption{Objective value 24 using constraints (\ref{iseven}), (\ref{yztree}), and (\ref{bobslatest})}
\label{3p20tree}       
\end{center}
\end{figure}

Spanning tree constraints help close the integrality gap on these long-path graphs because any spanning tree must contain $n-1$ edges, which will enforce $\sum_{e\in E} y_e + z_e \geq n-1$, where $n$ is the number of nodes in the graph.  For a relaxation on a graph that tries to save costs by employing frequent fractional $z$ variables, this constraint limits the amount that can be saved.  An optimal solution for a relaxation including constraint (\ref{yztree}) for the 3-path configuration of seven-edge paths is shown in figure~\ref{3p20tree}.

Spanning tree constraints (\ref{yztree}) did not contribute to smaller integrality gaps in our computational results of section \ref{seccomp} when added to the LP relaxation consisting of the subtour elimination constraints (or their compact equivalent), and the constraints in (\ref{bobslatest}) designed specifically to prevent the short half-$z$ path. 

\subsection{Removing Steiner Nodes}
\label{sec2c}

Removing Steiner nodes increase the effectiveness of constraints in (\ref{bobslatest}).
A graph with Steiner nodes $G = (V_d\cup V_s, E)$ can be transformed into a graph without Steiner nodes $G' = (V_d, E')$ by doing the following:

For each pair of nodes $i,j\in V_d$, if the shortest path from $i$ to $j$ in $G$ contains no other nodes in $V_d$, then add an edge connecting $i$ to $j$ to $E'$ with a cost equal to the cost of this shortest path; otherwise, do not add an edge from $i$ to $j$ to $E'$.

In all but one of our test problems (see section \ref{seccomp}), removing Steiner nodes resulted in fewer, not more, edges in the original instance.  That removing Steiner nodes often reduces the total edges in a graph was also observed by Corber\'an, Letchford, and Sanchis \cite{CLS}.

\section{Computational Results} 
\label{seccomp}

\begin{figure}[h]
\includegraphics[width=0.9\textwidth]{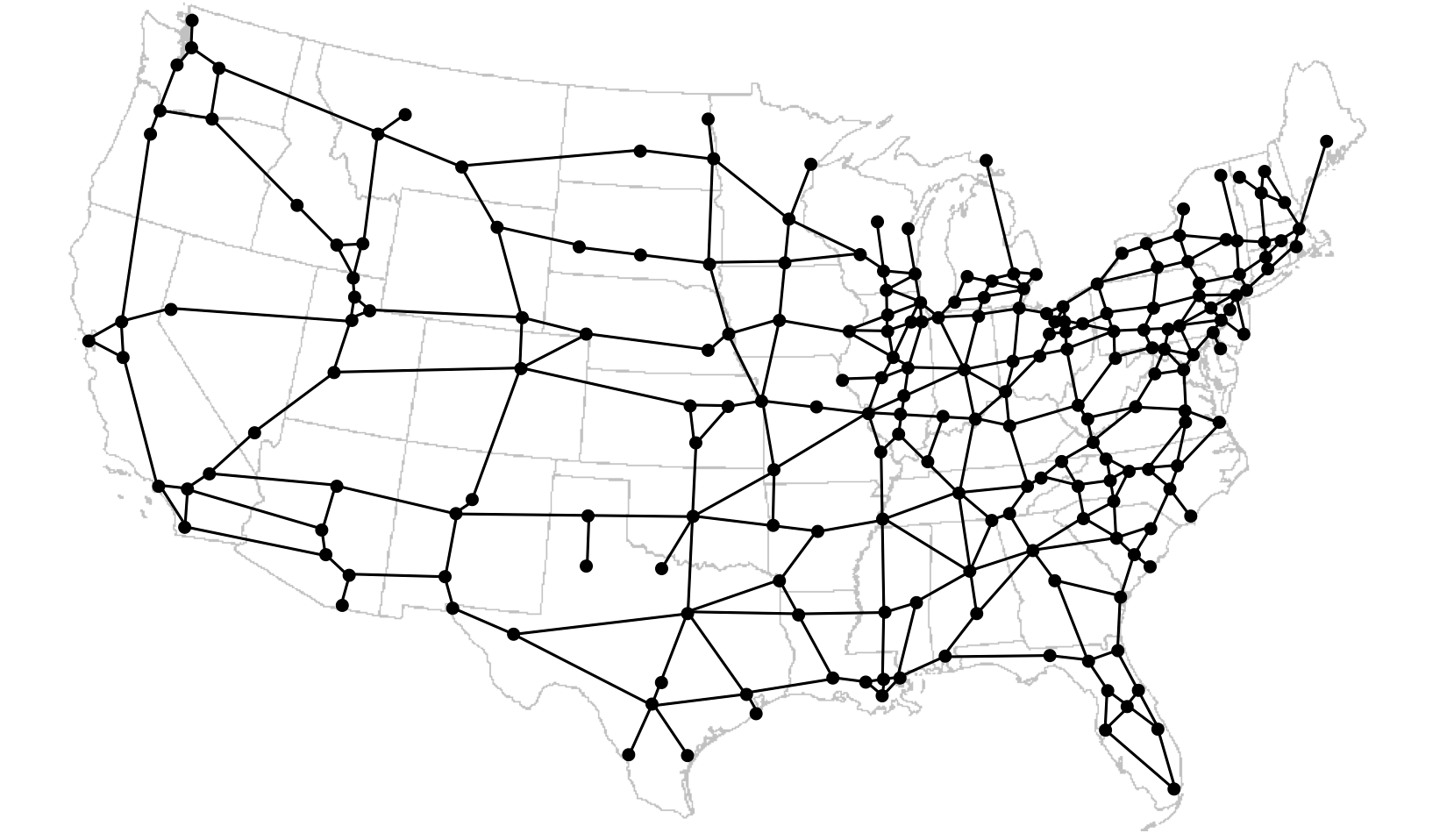}
\caption{Map of basic United States highway system}
\label{fig2}       
\end{figure}

\begin{table}[b]
\caption{GTSP instances}
\label{tabInst}       
\begin{tabular}{llcc}
\hline\noalign{\smallskip}
& & & Solution \\
name & desccription & destinations & (miles)  \\
\noalign{\smallskip}\hline\noalign{\smallskip}
dakota3path & 3-path configuarion in northern plains & 11 & 2682 \\
NFLcities & Cities with National Football League teams & 29 & 11050 \\
NWcities & Cities in the Northwest region & 43 & 8119 \\
CAPcities & 48 state capitals plus Washington D.C. & 49 & 14878 \\
AtoJcities & Cities beginning with letters from A to J & 101 & 17931 \\
ESTcities & Cities east of the Mississippi River & 139 & 13251 \\
MSAcities & Centers of 145 metropolitan statistical areas & 145 & 22720 \\
deg3cities & Cities in original graph with degree~$\geq 3$ & 171 & 18737 \\
NScities & Cities that Neil Simonetti has visited & 174 & 22127 \\
CtoWcities & Cities beginning with letters from C to W & 182 & 24389 \\
ALLcities & Entire graph & 216 & 26410 \\
\noalign{\smallskip}\hline
\end{tabular}
\end{table}

Our search for a reasonable sized data set based on the interstate highways of the United States led us to a text file uploaded by Sergiy Kolodyazhnyy on GitHub \cite{Serg}.  After a few errors were corrected and additions made, we had a highway network with 216 nodes and 358 edges, with a maximum degree node of seven (Indianapolis).  See figure~\ref{fig2}.  Data for this graph, and the cities used to create the instances in this section, may be found at https://ns.faculty.brynathyn.edu/interstate/

Instances were created from this map by choosing a subset of cities as destination nodes, and adding any cities along a shortest path between destinations as Steiner nodes.  Alternate versions of these instances were constructed by removing the Steiner nodes as indicated in section~\ref{sec2c}.  Table~\ref{tabInst} gives the basic information for several instances we used.  Table~\ref{tabCFN} shows the results from running the relaxation of the formulation from Cornu\'ejols et al. \cite{CFN}.  It should be noted that the solutions found by this relaxation were the same whether Steiner nodes were removed or not.

\begin{table}[h]
\caption{Relaxations from Cornu\'ejols et al. formulation \cite{CFN}}
\label{tabCFN}       
\begin{tabular}{lccc}
\hline\noalign{\smallskip}
& destinations & edges & integrality \\
name & (Steiner nodes) & (w/o Steiner) & gap (\%)  \\
\noalign{\smallskip}\hline\noalign{\smallskip}
dakota3path & 11 (0) & 12 (12) & 139 (5.18\%) \\
NFLcities & 29 (152) & 304 (135) & 35 (0.32\%) \\
NWcities & 43 (4) & 63 (59) & 12 (0.15\%) \\
CAPcities & 49 (132) & 301 (199) & 34 (0.23\%) \\
AtoJcities & 101 (95) & 326 (289) & 261.5 (1.45\%) \\
ESTcities & 139 (2) & 243 (240) & 61.4 (0.46\%) \\
MSAcities & 145 (63) & 348 (317) & 143 (0.63\%) \\
deg3cities & 171 (16) & 321 (305) & 70 (0.37\%) \\
NScities & 174 (29) & 341 (324) & 93.5 (0.42\%) \\
CtoWcities & 182 (30) & 353 (358) & 151 (0.62\%) \\
ALLcities & 216 (0) & 358 (358) & 274.8 (1.04\%) \\
\noalign{\smallskip}\hline
\end{tabular}
\end{table}

\begin{table} [h]
\caption{Relaxations from our additional constraints}
\label{tabUS}       
\begin{tabular}{lcccc}
\hline\noalign{\smallskip}
& integrality gap with & integrality gap w/o & best \% of gap closed \\
name & Steiner nodes (\%) & Steiner nodes (\%) & from formulation in \cite{CFN}  \\
\noalign{\smallskip}\hline\noalign{\smallskip}
dakota3path & - & 0 (0\%) & 100\% \\
NFLcities & 35 (0.32\%) & same as Steiner & 0\% \\
NWcities & 8 (0.10\%) & same as Steiner & 33.3\% \\
CAPcities & 34 (0.23\%) & same as Steiner & 0\% \\
AtoJcities & 228.5 (1.45\%) & same as Steiner & 12.6\% \\
ESTcities & 53.9 (0.46\%) & same as Steiner & 12.2\% \\
MSAcities & 114.5 (0.50\%) & 98.5 (0.43\%) & 31.1\% \\
deg3cities & 70 (0.37\%) & same as Steiner & 0\% \\
NScities & 48.5 (0.22\%) & same as Steiner & 48.1\% \\
CtoWcities & 103 (0.42\%) & 111.5 (0.46\%) & 31.8\% \\
ALLcities & - & 217.8 (0.82\%) & 20.7\% \\
\noalign{\smallskip}\hline
\end{tabular}
\end{table}

Running times on a 2.1GHz Xeon processor for all of the relaxations were under 10 seconds, while the running times to generate the integer solutions never exceeded five minutes.  We wish to point out that the value of the new formulation is not a faster running time, but the reduced integrality gap.

In this paper, the integrality gap refers to the difference in objective values between the program where integer constraints are enforced and the program where the integer constraints are relaxed.  The percentage is the gap size expressed as a percentage of the integer solution value.  This is different than the ratio definitions of integrality gap used in some other contexts. \cite{CV}

When our constraints were added, the spanning tree constraints (\ref{yztree}) were not  useful when (\ref{iseven}) and (\ref{bobslatest}) were present.  In most cases, removing Steiner nodes did not change the optimal values found by our relaxation.  In one case, the relaxation was better when the Steiner nodes were removed, and in one case, the relaxation was worse when the Steiner nodes were removed.  Table~\ref{tabUS} shows our results, where the last column indicates the percentage that our formulation closed of the gap left by the formulation of Cornu\'ejols et al. \cite{CFN}.

\begin{figure} [h]
\begin{center}
\begin{tikzpicture}
	\begin{axis}[
		xlabel=Percent of C-F-N integrality gap closed,
		ylabel=Percent of $x_e > 0$ edges also with $z_e > 0$]
	\addplot[scatter, only marks, point meta=explicit symbolic, scatter/classes={
			a={mark size=0.5pt},
			b={mark size=0.8pt},
			c={mark size=1pt},
			d={mark size=1.5pt},
			e={mark size=1.8pt},
			f={mark size=2pt},
			g={mark size=2.2pt}
		}] 
		table [meta=lab] {
		x      y      lab    
		0   0   b
		0 3.65  c
		33.33 20.83 c
		100 18.18 a
		12.6 8.41 d
		12.21 12.73 e
		48.1 10.99 f
		0 3.15 f
		20.7 13.01 g
		31.12 12.58 e
		26.16 9.95 f
	};
	\addplot[scatter, only marks, nodes near coords, nodes near coords style={font=\small}, point meta=explicit symbolic, mark size=0pt] 
		table [meta=lab] {
		x      y      lab    
		13 -1.1  NFLcities 
		13 3.1 CAPcities 
		46.0 19.73 NWcities 
		84 17.08 dakota3path  
		25.6 7.2 AtoJcities 
		12.21 10.53 ESTcities 
		59.5 9.95 NScities 
		13 1.5 deg3cities
		20.7 13.1 ALLcities 
		45.1 11.5 MSAcities
		41.5 8.7 CtoWcities
		70 0.0 {Size of dot is proportional}
		70 -1.0 {to number of destinations}
		70 4.0 {All instances have}
		70 2.7 {Steiner nodes removed}  
	};
	\end{axis}
\end{tikzpicture}
\end{center}
\caption{Scatter plot of integrality gap closure and percent of variables with $z_e > 0$}
\label{fig3}       
\end{figure}
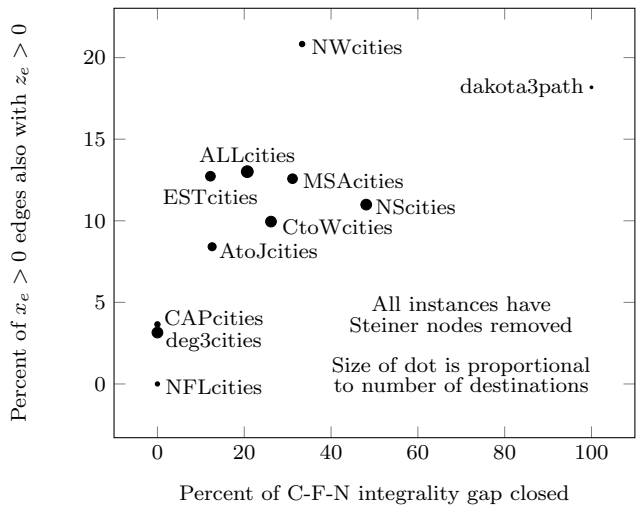

We noticed that in instances where the $z_e$ variables were rarely positive, our relaxation fared no better than that of Cornu\'ejols et al. But when the number of edges with values of $z_e > 0$ reached about 10\% of the total of edges where $x_e > 0$, we were able to shave anywhere from 10\% to almost 50\% of the gap left behind by Cornu\'ejols et al. (See figure~\ref{fig3})

\section{A Note Concerning T-joins} 
\label{Tjoinsec}
In section \ref{sec1}, we noted that the constraints of requiring a sum of variables to be even was unique and difficult, which
was a reason Naddef proposed his challenge, described in section \ref{mip}.  At first glance, these constraints may appear
to have the same type of structure as the T-join problem, but this is not the case, as seen in a book by Cook, et al. \cite{Tjoin}.

Let $G = (V,E)$ be an undirected graph and let $T \subset V$ with $|T|$ even.  
A T-join is a subgraph $H$ of $G$ where the set of all nodes of odd degree in $H$ is $T$.
The T-join polytope would therefore consist of all edge vectors $\bm{t}$ which indicate a T-join $H$.
The constraint for this polytope would be $t(\delta(v))$ is odd for all vertices $v$ in $T$ and even for all vertices $v$ not in $T$.
\begin{displaymath}
\begin{array}{lrlll}
&\frac{t(\delta(v))-1}{2} &\in& \mathbb{Z} & \mbox{for all $v \in T$}\\
\\
&\frac{t(\delta(v))}{2} &\in& \mathbb{Z} & \mbox{for all $v \not\in T$} \\
\end{array}
\end{displaymath}
While this appears to have the same issue as our GTSP formulation of section \ref{sec1}, the T-join problem can be
described very differently.  The common application of the T-join problem, used in solving the Chinese postman problem \cite{CPP}
assumes nonnegative costs within an IP formulation seeking a minimum cost T-join, and therefore
we only need the dominant polytope, which can be described as: 
\begin{displaymath}
\begin{array}{lrlll}
&t(\delta(S)) &\geq& 1 & \mbox{for any set $S$ where $|S \cap T|$ is odd} \\
\end{array}
\end{displaymath}




\end{document}